# BINDING OF HOLES TO MAGNETIC IMPURITIES IN A STRONGLY CORRELATED SYSTEM

D. POILBLANC, D. J. SCALAPINO [*], and W. HANKE [**]

*Laboratoire de Physique Quantique,*

*Université Paul Sabatier,*

*31062 Toulouse, France.*

## Abstract

The effect of a magnetic (S=1/2) impurity coupled to a 2D system of correlated electrons (described by the t–J model) is studied by exact diagonalisations. It is found that, if the exchange coupling of the impurity with the neighboring spins is ferromagnetic or weakly antiferromagnetic, an extra hole can form bound states of different spatial symmetries with the impurity extending to a few lattice spacings. The binding energy is maximum when the impurity is completely decoupled (vacancy) and vanishes for an antiferromagnetic coupling exceeding $\sim 0.3J$. Several peaks appear in the single hole spectral function below the lower edge of the quasiparticle band as signatures of the d-, s- and p-wave boundstates.

Typeset using REVTeX

The effect of doping the cuprate superconductors with Zn and Ni impurities have been explored in a variety of studies. These range from determining the depression of $T_c$ [1] to studying the temperature dependence of the penetration depth [2], conductivity, Knight shift and nuclear relaxation rates [3]. The picture that emerges is that at low concentrations both Zn and Ni impurities go into the planes Cu(2) sites and that Zn impurities are more damaging to the superconductivity state than Ni. If the Zn and Ni impurities maintain the nominal $Cu^{2+}$ charge, the $Zn^{2+}$ impurity would have a $(3d)^{10}$, S=0 configuration, while $Ni^{2+}$ would have a $(3d)^8$, S=1 configuration. Within this framework then the question which arises is why do the Zn impurities suppress superconductivity more strongly than a similar concentration of Ni impurities. One notion that would appear to play an important role is that if the pairing has $d_{x^2-y^2}$ orbital symmetry rather than the usual s-wave symmetry one can at least understand why Ni is not overwhelmingly more effective in suppressing superconductivity than Zn. However, this still leaves the fact that Zn is more effective than Ni to be explained.

Here it is natural to look for a qualitative understanding by recognizing that the pure $CuO_2$ layer represents a highly correlated electron system with strong short range antiferromagnetic (AF) correlations between the Cu sites. A Zn impurity with S=0 acts like an inert site, eliminating the antiferromagnetic exchange couplings to its near neighbor Cu sites and disrupting the local AF correlations. In addition the neighboring copper sites may develop local moments. In a previous study [4] we modeled this using a t–J model with an inert site. This site had no exchange coupling or electron transfer terms. We found that a hole doped into this system experiences an extended effective impurity potential which depending upon the magnitude J/t of the host could form bound d,s and p states.

Here we consider a similar approximate characterization of the Ni impurity problem. For this case, we allow the impurity to have a spin (which for simplicity we take to be S=1/2) and couple it to its near neighbors via an exchange coupling $J'$. We continue to exclude the electron transfers onto the impurity site. The Hamiltonian reads,

$$H = J \sum_{t=x,y} \sum_{i \neq o, -t} (\mathbf{S_i} \cdot \mathbf{S_{i+t}} - \frac{1}{4} n_i n_{i+t}) + t \sum_{t=x,y} \sum_{i \neq o, -t, s} P_G (c_{i,s}^\dagger c_{i+t,s} + h.c.) P_G$$
$$+ J' \sum_{t=\pm x, \pm y} (\mathbf{S_o} \cdot \mathbf{S_t} - \frac{1}{4} n_t). \qquad (1)$$

A standard t–J model is used to describe the strongly correlated host, $\mathbf{S_i}$ and $c_{i,s}^\dagger$ being electron spin and creation operators. The Gutwiller projector $P_G$ excludes higher energy configurations with doubly occupied sites. Note that the 4 bonds connected to the impurity are excluded in the first two terms. Hence, holes in the AF feel a bare *hard-core repulsive interaction* at the impurity site. The last term of (1) corresponds to the coupling (ferro- or antiferromagnetic) of the impurity to the neighboring spins of the correlated system. No charge fluctuations are present at the impurity site (because of the infinite on-site hole repulsion) so that $\langle n_o \rangle = 1$. Hereafter the impurity spin is set to $S = 1/2$ but we believe (1) can also qualitatively describe the effects of a higher spin impurity (as e.g. a S=1 Ni impurity). Although (1) is similar to the simple Kondo impurity model, the strongly correlated nature of the host is here of central importance and leads to qualitatively new physics.

When the impurity coupling $J'$ vanishes (1) reduces to the vacancy model i.e. of a spinless impurity in a strongly correlated system, (e.g. Zn in cuprate superconductors). As mentioned above, although a vacant site cannot accomodate holes, the local deformation of the antiferromagnet [5] produces in that case a dynamical attractive potential for an extra hole [4] that leads to bound states. In the case of a *magnetic* impurity we shall consider instead $J' \neq 0$ leading to non-conservation of the total spin $\sum_{i \neq o} \mathbf{S_i}$ of the host. The purpose of this paper is to study the effect of this coupling on the binding energies and on the dynamical properties of an extra hole. Exact diagonalizations by a Lanczos algorithm of small clusters (typically of $N = \sqrt{20} \times \sqrt{20}$ sites) of Hamiltonian (1) at half-filling or with a single hole are performed to calculate GS energies and dynamical correlations. Full use of rotation symmetry is necessary to partially block-diagonalize the hamiltonian matrix [6].

First, it is useful to compare the single hole GS energy in the presence of the impurity to the single hole GS energy in the pure system [7]. If negative, the difference of these quantities, $\Delta_B = (E^{imp}_{1h,0} - E^{imp}_{0h,0}) - (E^{pure}_{1h,0} - E^{pure}_{0h,0})$, where the superscript (subscript) refers to the presence or absence of an impurity (a hole) in the system, can be interpreted as the binding energy (negative) of the single hole to the impurity [4]. In other words, if we consider two clusters of size N, binding occurs when the energy $E^{imp}_{1h,0} + E^{pure}_{0h,0}$ corresponding to the impurity and the hole confined in the same cluster is lower than the energy $E^{imp}_{0h,0} + E^{pure}_{1h,0}$ of the hole and the impurity in separate clusters. Similar techniques were recently used to establish the existence of an effective attraction between holes doped in the AF [8]. The binding energy vs $J'/J$ has been calculated for various spatial symmetries of the single hole GS and results are shown in fig. 1 for a physical value of $J/t = 0.5$. For a small coupling $J'$, d-, s- and p-wave boundstates coexist ($\Delta_B < 0$), the d-wave being the most stable. The largest binding energies are found when the impurity is magnetically decoupled from the spins of the host ($J' = 0$) i.e. when the impurity behaves as a vacancy [4]. The sign of the magnetic coupling $J'$ seems to be crutial. An increasing *antiferromagnetic* coupling leads to a rapid reduction of the binding energies. Typically, a small coupling around $J'/J \sim 0.3$ is sufficient to prevent any binding. This can be easily understood; indeed, when e.g. $J'$ equals exactly $J$ the localized impurity has no effect on the spin background, and the hole feels only a hard-core repulsive potential. The strong sensitivity to a positive $J'$ is then evidence that the binding occurs through many-body effects i.e. through a local deformation of the surrounding spins. A ferromagnetic coupling has a much weaker effect on the binding energy. It should be noticed that, for a sufficiently large ferromagnetic coupling the s-wave bound-state becomes the more stable.

The particular cusp-like behavior at $J' = 0$ of the binding energies in fig. 1 deserves special attention. When $J' \to 0$, the impurity spin becomes free and the total spin S=1/2 as well as the z-component $S_z = \frac{1}{2}\sigma$ of the remaining $N - 1$ spins (at half-filling) is then conserved. The corresponding degenerate wavefunctions $|\Psi^{imp}_{0,\sigma}\rangle$ ($\sigma = \pm 1$) have been calculated previously [4]. The singlet and triplet GS of the full N-spin system are then tensorial products,

$$\lim_{J' \to 0} |\Psi_{0h,0}^{imp}\rangle = \frac{1}{\sqrt{2}} \{(c_{\mathbf{o},-\sigma}^\dagger |0\rangle) \bigotimes |\Psi_{0,\sigma}^{imp}\rangle \pm (c_{\mathbf{o},\sigma}^\dagger |0\rangle) \bigotimes |\Psi_{0,-\sigma}^{imp}\rangle\}. \qquad (2)$$

The exact degeneracy at $J' = 0$ is actually lifted by an infinitesimally small $J'$. Hence, a crossing between the triplet GS energy (for $J' < 0$) and the singlet GS energy (for $J' > 0$) occurs at $J' = 0$. Such a level crossing is responsible for the cusp-like singularities in fig. 1.

Useful informations on the single hole bound state wavefunction can be obtained from the local density of state,

$$N_{\mathbf{ii}}^\sigma(\omega) = -\frac{1}{\pi} Im \{\langle \Psi_{0h,0}^{imp}| c_{\mathbf{i},\sigma}^\dagger \frac{1}{\omega + i 0^+ - H + E_{0h,0}^{imp}} c_{\mathbf{i},\sigma} |\Psi_{0h,0}^{imp}\rangle\}, \qquad (3)$$

where $\mathbf{i} \neq \mathbf{o}$ and $E_{0h,0}^{imp}$ is the GS energy of a single impurity at half-filling. $N_{\mathbf{ii}}^\sigma(\omega)$ is spin ($\sigma$) independent since the z-component of the total spin $S_z$ equals 0 (N even). Moreover it also fulfills the sum rule $\int_{-\infty}^{+\infty} N_{\mathbf{ii}}^\sigma(\omega) d\omega = \frac{1}{2}$ as in the pure system.

As this stage it is interesting to investigate the limit $J' \to 0$. Using Eq. (2) it is clear that the spectral function splits into two terms,

$$\lim_{J' \to 0} N_{\mathbf{ii}}^\sigma(\omega) = N_{\mathbf{ii},vac}^{\sigma,\sigma}(\omega) + N_{\mathbf{ii},vac}^{\sigma,-\sigma}(\omega). \qquad (4)$$

The first (second) term corresponds to preparing the initial single hole state with a total spin component $S_z = 0$ ($|S_z| = 1$) by removing a spin $\sigma$ from the $|\Psi_{0,\sigma}^{imp}\rangle$ ($|\Psi_{0,-\sigma}^{imp}\rangle$) GS. Hence, $N_{\mathbf{ii},vac}^{\sigma,\sigma}$ contains both singlet and triplet excitations of the N-2 remaining spin system (calculated in Ref. [4]) while only higher energy triplet excitations appear in the spectral decomposition of $N_{\mathbf{ii},vac}^{\sigma,-\sigma}(\omega)$. When $J'$ is switched on the two terms mix since the z-component of the total spin of the N-2 spin sub-system starts to fluctuate.

As in Ref. [4] the local density (3) is calculated by decomposing the hole creation operator $c_{\mathbf{i},\sigma}$ into its symmetric components; by applying all point group elements $\mathcal{S}$ (with the origin at the impurity site $\mathbf{o}$) to a given site $\mathbf{i}$ one generates a set of equivalent sites $\mathbf{i}_\mathcal{S} = \mathcal{S}(\mathbf{i})$ [9]. One can then construct the symmetric operators,

$$c_{\mathbf{i},\sigma}^\alpha = (\mathcal{N}_{\mathbf{i}}^\alpha)^{-1/2} \sum_\mathcal{S} c_\mathcal{S}^\alpha(|\mathbf{i}|) c_{\mathbf{i}_\mathcal{S},\sigma}, \qquad (5)$$

where $\alpha$ labels the irreducible point group representations (which depend on the site location), $c_S^\alpha(|\mathbf{i}|) = 0, \pm 1$ are the corresponding characters and $\mathcal{N}_\mathbf{i}^\alpha$ are normalization factors satisfying $\sum_\alpha (\mathcal{N}_\mathbf{i}^\alpha)^{-1} = 1$. The relations (5) can easily be inverted and the expressions for $c_{\mathbf{i},\sigma}$ substituted into (3). The calculation of the density of states is then made separately in each symmetry sector and the various components added afterwards with the appropriate weights.

The local density of states on the neighboring sites to the impurity is shown in fig. 2. At small $J'$, various peaks can be seen at the bound state energies $\Delta_B^\alpha$ (see fig. 1) below the band edge of the pure system [10]. Although we cannot completely exclude the possibility of a local shift of the chemical potential in the vicinity of the impurity, we would rather interpret these peaks as true $\delta$-peaks characteristic of real bound states (broadening in the figures are artificially caused by a small imaginary part in the frequency). We then expect the local density of states below the Fermi level to be of the form $N_{\mathbf{ii}}^\sigma(\omega) \sim \sum_\alpha Z_\mathbf{i}^\alpha \delta(\omega - \Delta_B^\alpha)$ where the sum runs over a *finite* number of states. True bound states have finite weights $Z_\mathbf{i}^\alpha$ in the thermodynamic limit (in contrast to the case of resonances) and are separated from the continuum. A moderate AF coupling $J'$ is sufficient to push these states back into the continuum as seen in fig. 2c.

Figs. 3a and 3b show that the various bound states have significant weights at larger distances from the impurity (here at distance $\sqrt{5}$ [9]). Hence, as in the vacancy case, the hole-impurity bound pair is extended in space on a few lattice sites in contrast to the simple case of an attractive scattering potential in a non–interacting tight–binding electron gas [11]. We also observed that, if the impurity sits let say on the A sublattice the spectral weights $Z_\mathbf{i}^\alpha$ on the B sites is generally quite small although the bound state wave functions itself spread on these sites. This effect can be attributed to relaxation or retardation effects as also recently observed in bound pair of holes [13]. In addition we note that, on finite clusters, the total weight $\sum_{\mathbf{i} \neq \mathbf{o}} Z_\mathbf{i}^\alpha$ (integrated over space) of a given boundstate is in general smaller than 1 (typically in the range 0.3 – 0.7) while Friedel's sum rule would predict an exact integer number of states to be pulled out of the continuum [12]. Whether this violation also holds in the thermodynamic limit needs certainly to be further investigated. Lastly, we observe by comparing fig. 3a, 3b and 3c with increasing $J'$ the appearence of a resonance around $\omega \sim -0.17$ which corresponds almost exactly to the top of the quasi–particle band [10]. This resonance could be interpreted as a reminiscence of the bound state existing in a tight binding model in the presence of a strong repulsive scattering potential [11]. However, in a strongly correlated system, multi–spin wave scattering leads to incoherent weight above the quasi–particle band [10] and hence gives a finite life time to the bound state

We finish this paper by a few comments on the relevance of hamiltonian (1) to the case of Ni impurities in high-$T_c$ cuprate superconductors. Clearly the possible hybridization of the Ni orbitals with the surrounding oxygen orbitals has been neglected here. This could be taken into account in our model by including a small hopping $t'$ between the impurity and the neighboring sites. One can also generalize (1) to the case of a spin 1 impurity. Such extensions are left for future work.

In summary, carrying out exact Lanczos diagonalizations of a simple model of a magnetic impurity imbeded in a strongly correlated host, the binding energy and local site spectral weights of an added hole are calculated. We find that if the exchange coupling $J'$ of the impurity with its near neighbor spins is ferromagnetic or weakly ($J'/J \ll 1$) AF, the added hole forms d,s and p bound states just as in the case of an inert impurity. However, it is clear that if $J' = J$, then the spin correlation would be unchanged and the impurity would simply act as a hard core potential with no bound states. We find that even for a modest $J'/J$ coupling ($> 0.3$) there are no bound states. Then, within this model, we picture the Ni impurities as maintaining the local AF correlations and hence acting as weaker scattering centers than the Zn impurities which destroy the local Cu spin correlations.

### Acknowledgment

DJS and DP acknowledge support from the National Science Foundation under grant

DMR92-25027 and from the EEC Human Capital and Mobility program under grant CHRX-CT93-0332 respectively. We also thank IDRIS, Orsay, France for allocation of CPU time on the CRAY-C98 supercomputor.

# REFERENCES

\* Permanent address: Department of Physics, University of California at Santa Barbara, Santa Barbara, CA 93106

\*\* Permanent address: Institute for Physics, University of Würzburg, D-8700 Würzburg, Germany.

**FIG. 1**

Binding energy of the d, s and p-wave bound states on a 20-site cluster as a function of the impurity exchange coupling to the neighboring spins. The ratio $J/t$ is fixed to 0.5.

**FIG. 2**

Local density of state on sites at distances $R_i = 1$ from the impurity site on a $\sqrt{20} \times \sqrt{20}$ cluster for fixed $J/t = 0.5$. The energy of the d-wave bound state bound in the case of a *vacancy* ($J' = 0$) is indicated by a thin dashed line (see Ref. [4]). The lower edge of the (upper) Hubbard band of the pure system is shown as a reference by a thicker dashed line. Figs. (a), (b) and (c) correspond to different impurity couplings, $J'/J = -0.2$, 0.1 and 1 respectively.

**FIG. 3**

Same as fig. 2 but at distances $R_i = \sqrt{5}$ from the impurity site.

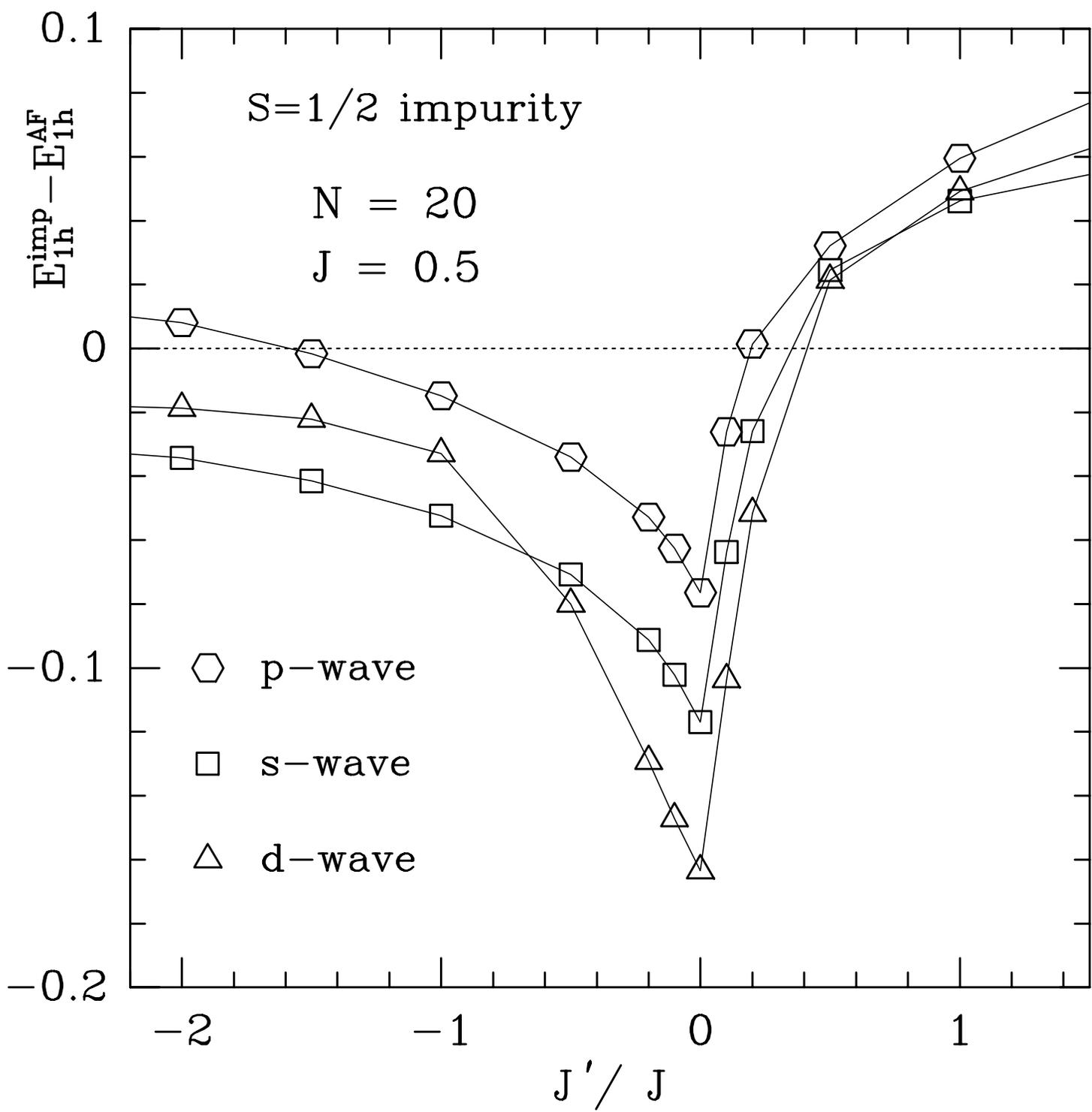

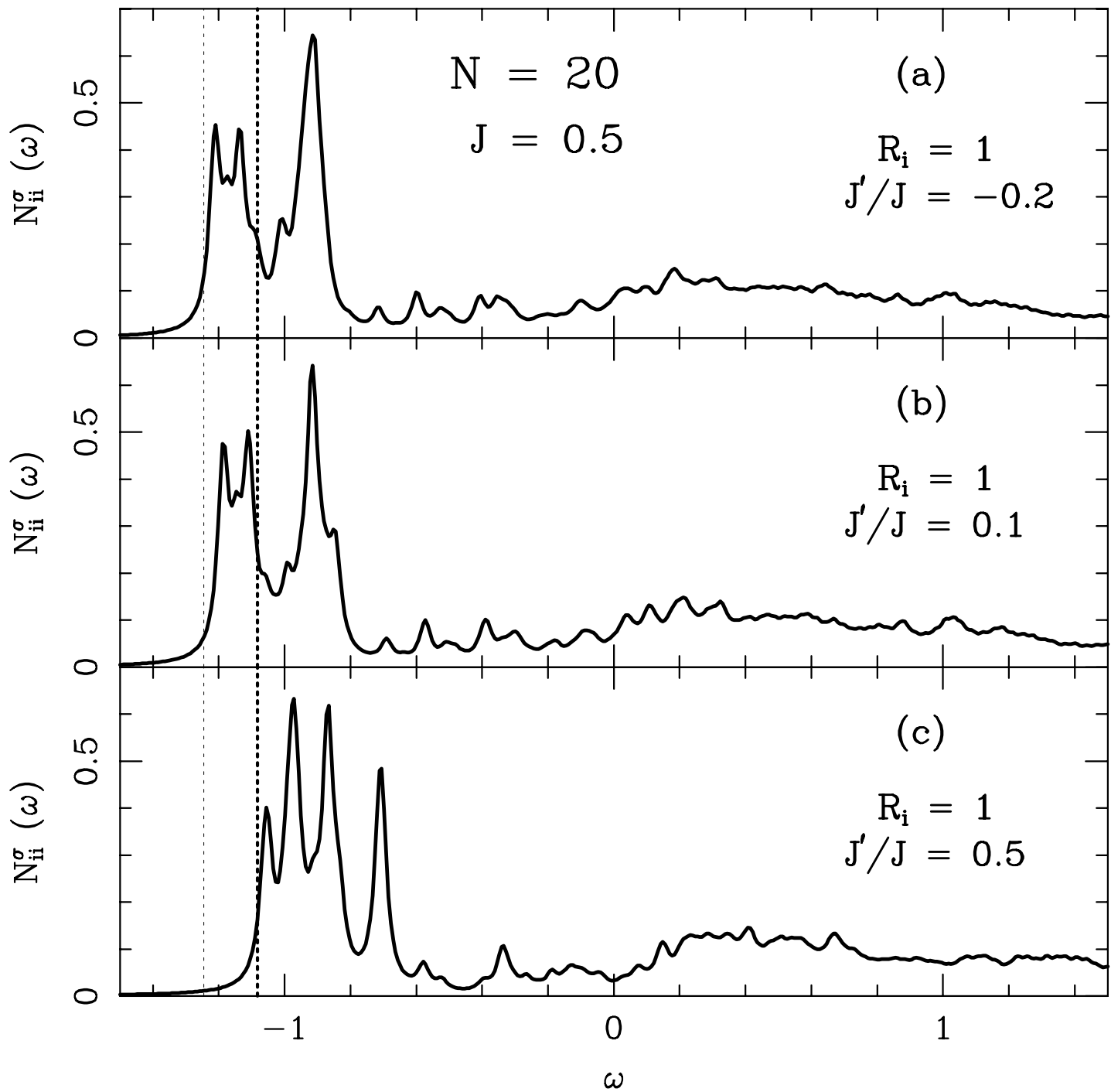

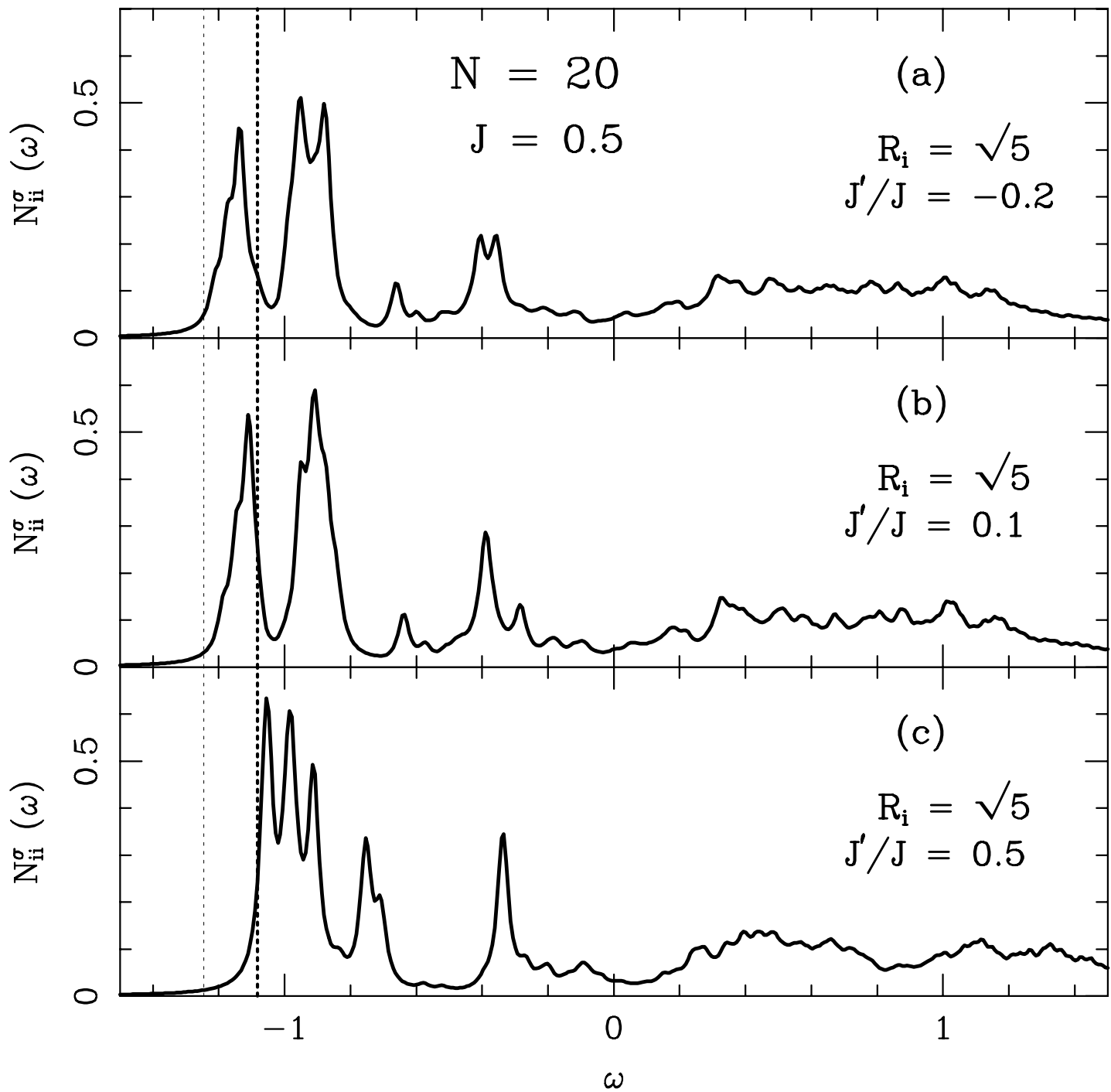